\documentclass[12pt]{iopart}
\usepackage{graphicx}

\begin{document}

\title[Suppression of the superconducting energy gap in IJJs of BSCCO single crystals]{Suppression of the superconducting energy gap in intrinsic Josephson junctions of $\mathbf{Bi_2Sr_2CaCu_2O_{8+\delta}}$ single crystals}

\author{L. X. You\dag\ddag, P. H. Wu\dag, J. Chen\dag, K. Kajiki\S, S. Watauchi\S\ and I. Tanaka\S}
\address{\dag\ Research Institute of Superconductor Electronics (RISE), Department of Electronic Science and Engineering, Nanjing University, Nanjing 210093, China}

\address{\ddag\ Quantum Device Physics Laboratory, Department of Microtechnology and
Nanoscience, Chalmers University of Technology, SE-412 96
G\"oteborg, Sweden}

\address{\S\ Center for Crystal Science and Technology, University of Yamanashi, Miyamae 7, Kofu, Yamanashi 400-8511, Japan}

\ead{lixing@mc2.chalmers.se}

\begin{abstract}
We have observed back-bending structures at high bias current in
the current-voltage curves of intrinsic Josephson junctions. These
structures may be caused by nonequilibrium quasiparticle injection
and/or Joule heating. The energy gap suppression varies
considerably with temperature. Different levels of the suppression
are observed when the same level of current passes through top
electrodes of different sizes. Another effect which is seen and
discussed, is a super-current ``reentrance'' of a single intrinsic
Josephson junction with high bias current.
\end{abstract}


\maketitle

\section{Introduction}

It is difficult to fabricate Josephson tunnel junctions in high
temperature superconductors (HTS), due to the small coherence
length of HTS (1-2 nm in the $ab$ plane and about 0.2 nm along the
$c$-axis). However, the natural layered structure in HTS along the
$c$-axis forms tunnelling coupling between adjacent
superconducting Cu-O layers with some non-superconducting layers
in between acting as the potential barrier (for example, in
$\mathrm{Bi_2Sr_2CaCu_2O_{8+\delta}}$(BSCCO), there are Bi-O and
Sr-O layers) \cite{Kleiner:PRL92,Oya:JJAP92}. Each tunnelling
junction is about 1.5 nm thick, implying that a thin BSCCO single
crystal with a thickness of 150 nm contains 100 intrinsic
Josephson junctions (IJJs) in series. The brush-like quasiparticle
branch structure in the current-voltage ($I-V$) characteristics is
a typical sign of IJJs. However, a suppression of the
superconducting energy gap is often observed at a high bias
current density. With all the junctions biased into their
resistive states, the last quasiparticle branch has two possible
appearances. One is an ordinary quasiparticle curve with positive
dynamic resistance
\cite{Winkler:SUST99,Irie:PhysicaC02,Kitamura:PRB02}; the other is
a back-bending structure with negative dynamic resistance
\cite{Prusseit:PhysicaC97,Sakai:PhysicaC97,Suzuki:PhysicaC97,Yurgens:APL97}.
This makes it difficult to obtain the real energy gap from $I-V$
curves of IJJs.

\par
We will present our observation of the energy gap suppression in
IJJs under high bias current density. We discuss possible
mechanism of the energy gap suppression and the temperature
dependence. Two different top electrode sizes allow injection of
different current densities at the same bias current. We observed
clearly different energy gap suppression in IJJs. A novel
super-current reentrance phenomena was also observed and will be
discussed.

\section{Sample fabrication}

\begin{figure}[tb]
\begin{center}
\includegraphics[width=.6\textwidth]{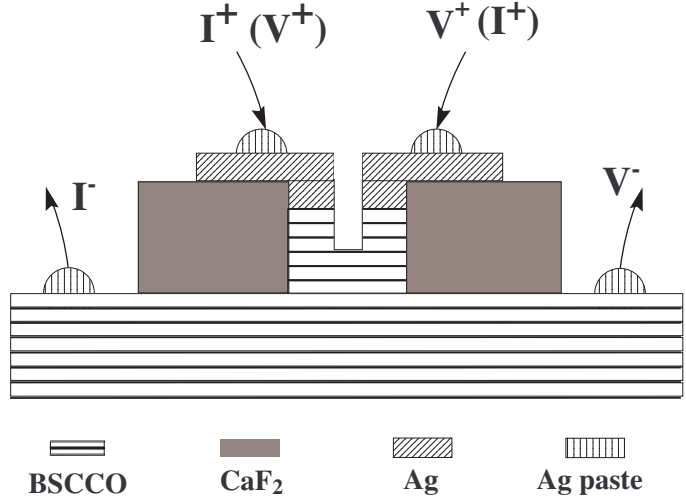}
\end{center}
\caption{Schematic view of mesa structured IJJs with four-terminal
arrangement} \label{mesa}
\end{figure}

BSCCO single crystals with critical temperature $T_c \approx 90$\
K were grown using the travelling solvent floating zone (TSFZ)
method. We fixed a piece of single crystal BSCCO on a Si substrate
with polyimide, and then cleaved it to reveal a fresh surface. A
thin-film silver layer was evaporated onto this surface. A square
mesa with an area of $\mathrm{8 \times 8\ \mu m^2}$ or $\mathrm{16
\times 16\ \mu m^2}$ was formed using conventional
photolithography and argon-ion etching. The number of junctions in
the mesa was controlled by the etching rate and time. A layer of
$\mathrm{CaF_2}$ was evaporated to isolate the mesa from the base
single crystal right after the etching step. Ultrasonic rinsing in
acetone was used to remove the photoresist and the
$\mathrm{CaF_2}$ layer on it. The mesa with the silver thin film
was subsequently covered with a second silver layer by
evaporation. Two separate top electrodes were formed on the mesa
by another photolithography and argon ion etching step. Another
two base electrodes can be easily formed on the base of the BSCCO
chip for four-terminal measurements. The schematic diagram is
shown in Figure \ref{mesa}. Note that the junctions in the bottom
of the U-shaped mesa are the effective junctions in four-terminal
measurements, whereas the junctions formed beneath the electrodes
act as a part of the contact resistance (they will be referred to
as `electrode junctions' in what follows). A detailed discussion
of the IJJs fabrication technology, the properties and the control
of the number of junction was published elsewhere
\cite{You:SUST03,You:JJAP04,You:PhysicaC04}. With this fabrication
method, we can realize mesa-structured IJJs with any number of
junction and eliminate the effect of the surface junction
\cite{Doh:PRB00}.

\section{Measurements and discussion}

\begin{figure}[tb]
\begin{center}
\includegraphics[width=.6\textwidth]{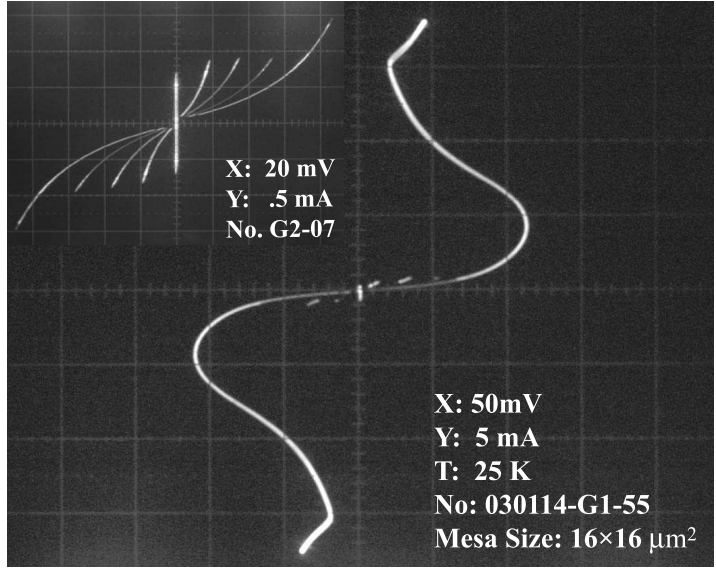}
\end{center}
\caption{Typical $I-V$ curves of mesa-structured IJJs containing
four junctions. The inset shows similar critical currents of the
four junctions.} \label{IV1}
\end{figure}

All the measurements were carried out in a cryocooler with a
minimal sample temperature of 14\ K. Figure \ref{IV1} shows $I-V$
curves of a sample containing four effective junctions. The size
of the mesa in the $ab$ plane is $\mathrm{16 \times 16\ \mu m^2}$.
Two top electrodes with the same size of $\mathrm{6 \times 16\ \mu
m^2}$ are separated by a trench of $\mathrm{4 \times 16\ \mu m^2}$
formed by the second ion etching. The fabrication parameters are
chosen in such a way that we have in the sample 4 effective
junctions and 3 electrode junctions. The $I-V$ curves in the inset
of Figure \ref{IV1} show clearly four branches with similar
critical currents of $I_c \sim 0.8$\ mA. At higher bias current,
shown in Figure \ref{IV1}, we observed the well-known back-bending
structure of $I-V$ curve with $V_{gmax}=27$\ mV and
$V_{gmin}=6.8$\ mV, where $V_{gmax}$ and $V_{gmin}$ denote the
average maximal and minimal voltages in the back-bending structure
per each junction separately.

\par
Tunnelling spectroscopy experiments and three-probe measurements
of the $I-V$ characteristics of ultra small mesa-structured IJJs
usually gives a typical value of the superconducting energy gap
$\Delta \sim 25 meV$
\cite{Kitamura:PRB02,Liu:PRB94,Vedeneev:PRB94,Renner:PRB95,Mueller:FPSJD02}.
The obtained $V_{gmax}$ is obviously less than $2\Delta/e$ (54\%
of $2\Delta/e$) and $V_{gmin}$ is only 14\% of $2\Delta/e$, which
illustrate a great suppression of the energy gap under high
current density injection.

\par
\begin{figure}[tb]
\begin{center}
\includegraphics[width=.6\textwidth]{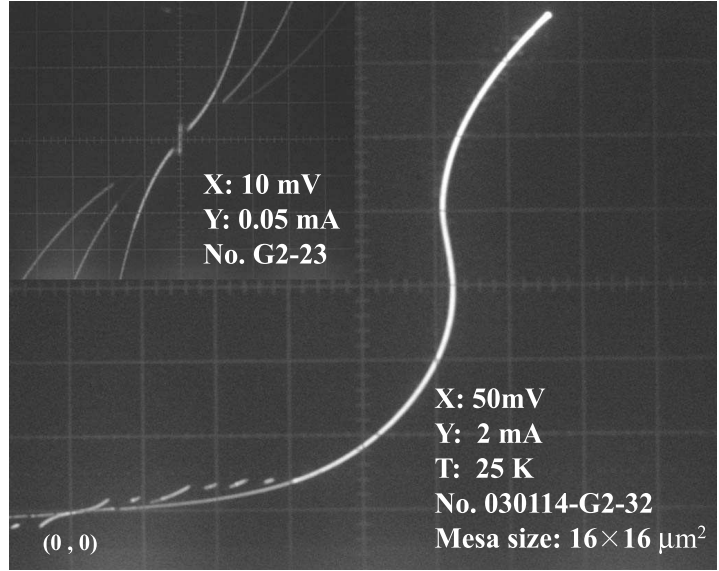}
\end{center}
\caption{Three-terminal $I-V$ curves of the sample sharing a top
electrode for both current bias and voltage measurement. The inset
shows the existence of a surface junction with a small critical
current} \label{IV-3T}
\end{figure}

For comparison, three-terminal measurements were also carried out
for the sample sharing a same top electrode for both current bias
and voltage measurement. Shown in Figure \ref{IV-3T} are the
three-terminal $I-V$ curves of the sample with 7 quasiparticle
branches, which represents the number of junctions in the mesa.
Due to the difference of the critical current, the branches are
divided into two groups, one group with $I_c= 0.3$\ mA corresponds
to the electrode junctions in four-terminal measurements; the
other group with $I_c= 0.8$\ mA corresponds to the four effective
junctions measured in four-terminal method. The magnified figure
near the origin shows a junction with a quite small critical
current of $I_c= 0.02$\ mA (see the inset of Figure \ref{IV-3T}),
which is the surface junction with suppressed superconductivity
formed in the interface between the silver layer and the BSCCO
mesa \cite{Doh:PRB00}. With higher bias current, we also observe a
quasiparticle branch with a slight back-bending structure. Because
of the inconsistence of those junctions in series in
three-terminal measurements, it is difficult for us to evaluate
the energy gap quantitatively of each junction. Further analysis
and discussions of the suppression of the energy gap will be
concentrated on the results of the four-terminal measurements.

\par
\begin{figure}[tb]
\begin{center}
\includegraphics[width=.6\textwidth]{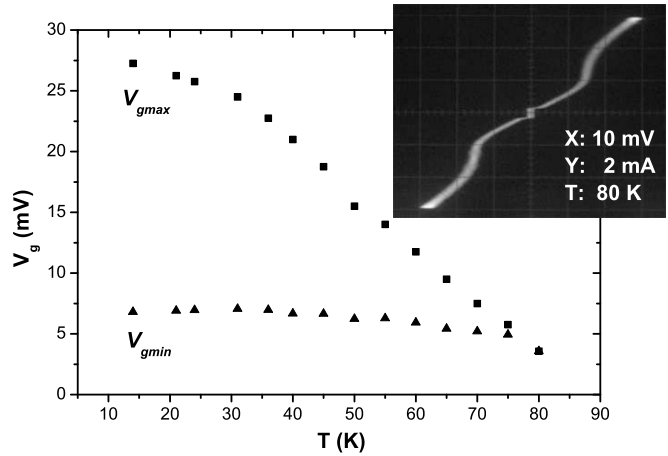}
\end{center}
\caption{Temperature dependence of $V_{gmax}$ and $V_{gmin}$ for a
4-junction IJJs. The inset shows the $I-V$ curve of the sample at
T=80 K} \label{Vg-T}
\end{figure}

To probe the relation between the energy gap suppression and
temperature, the dependence of $V_{gmax}$ and $V_{gmin}$ on
temperature was measured and is shown in Figure \ref{Vg-T}.
$V_{gmax}$ decreases quickly with increasing temperature; whereas,
$V_{gmin}$ remains relatively temperature independence with a
slight decrease when the temperature is near $T_c$. When T=80K,
$V_{gmax}=V_{gmin}$, and an abrupt quasiparticle curve was
observed instead of a back-bending structure (shown in the inset
of Figure \ref{Vg-T}). Generally, the energy gap voltage
$2\Delta/e$ decreases with increasing temperature, so $V_{gmax}$
decreases with a similar trend. On the other hand, the measurement
results show that the bias current for $V_{gmin}$ decreases due to
the increase of temperature. As a result, the Joule heating effect
and the nonequilibrium effect are reduced, making the back-bending
structure shrinking. Finally when the temperature is high enough
we have $V_{gmax}=V_{gmin}$, and the $I-V$ curve turns into an
abrupt quasiparticle curve.

\par
Although quite a few paper have discussed the suppression of the
energy gap and attributed to a quasiparticle injection
nonequilibrium effect and a Joule-heating effect, not many
detailed analysis are available
\cite{Prusseit:PhysicaC97,Sakai:PhysicaC97,Suzuki:PhysicaC97,Yurgens:APL97}.
To further address this problem, we will first look at the
possible quasiparticle injection in IJJs. Concerning the $c$-axis
quasiparticle transport, we remark that the effect of the current
injection depends strongly on the transmission and energy
relaxation of quasiparticles along the $c$-axis, and they are
therefore sensitive to the inter-planar inelastic scattering
mechanism in addition to the in-plane quasiparticle recombination.
Since the $c$-axis dimension of the mesa is much smaller than the
lateral dimensions, the effect of quasiparticle injection should
be primarily determined by the magnitude of the $c$-axis
quasiparticle relaxation length $\delta_n^c$, even though the
in-plane recombination time of excess quasiparticles can be
relatively long due to the existence of nodes in the pairing
potential \cite{Feenstra:PRL97}.

\par
$\delta_n^c$ may be expressed as $\delta_n^c=\sqrt{D_n^c \tau
_Q}$, where $D_n^c=v_F l_{tr}^c/3$ is the charge diffusion
coefficient along $c$-axis ($v_F=2 \times 10^5\ m/s$ and $l_{tr}^c
\sim 1\ nm$ are the Fermi velocity and transport mean-free path
along the $c$-axis respectively) \cite{Fu:PRB02}. It is known that
the characteristic quasiparticle relaxation time $\tau_Q$ follows
the relation \cite{Schmidt:Book97}
$$ \tau_Q(T) \approx \frac{4\tau_E k_B T_c}{\pi \Delta (T)}$$
where $\tau_E$ is the inelastic electron-phonon scattering time,
and $\Delta(T)=\Delta_0(1-T/T_c)^v$, with $\Delta_0$ being the
zero-temperature superconducting gap and $v$ the order parameter
critical exponent. For $T/T_c\ll 1$, with typical $ \tau_E \sim
10^{-11} s$ for a cuprate and energy gap $\Delta = 25 meV$ for
BSCCO, we have $\tau_Q \sim 4 ps$, and then deduce $\delta_n^c
\sim 16 nm$, which is much larger than the thickness of the Cu-O
layers. The estimate shows that the quasiparticles can easily pass
through the thin Cu-O layers of 3 \AA\ thick, then be injected
into the junctions along the $c$-axis and influence their
properties. Recent experiments estimated $\tau_Q$ to be about of
100 ps \cite{Rother:PRB03}. This would cause an even bigger
$c$-axis quasiparticle relaxation length and stronger
quasiparticle injection effect. The quasi-particle injection
effect may cause nonequilibrium gap suppression. This has been
discussed both theoretically and experimentally for tunnelling
junctions made of low temperature superconductors
\cite{Owen:PRL72,Parker:PRL72,Chang:JLTP78,Yeh:PRB78,Winkler:PS85,Dynes:PRL77}.
A similar back-bending structure was also observed in some of
their experiments \cite{Yeh:PRB78,Winkler:PS85}.
\par
On the other hand, all the high temperature superconductors have a
comparatively low thermal conductivity, which makes them prone to
local overheating. Both estimates and experimental results have
shown that the temperature of a mesa may increase by several tens
of degrees when the bias current is high
\cite{Yurgens:SUST00,Thomas:Arxiv00,Yurgens:Arxiv03}. Concerning
the sample with $I-V$ curves shown in Figure
\ref{IV1}\&\ref{IV-3T}, the thermal dissipation power for the
whole junction stack at the bias current of 4 mA (the bias current
for $V_{gmax}$) can be estimated to be about 1 mW from Figure
\ref{IV-3T}. If we adopt a typical value for the thermal
conductance of BSCCO (40 $\sim$ 70 K/mW) \cite{Yurgens:Arxiv03},
the possible temperature increase would be over 40 K. Such a
serious Joule heating effect will, of course, cause a back-bending
structure in the $I-V$ curve. Recently, a short-pulse measurement
technique was adopted to minimize the effect of overheating in
IJJs, and the back-bending was indeed removed.
\cite{Fenton:APL02,Hamatani:PhysicaC03}.

\par
Since both Joule heating and nonequilibrium effects originate from
a high quasiparticle current density, it is difficult to
distinguish the two effects and hard to say which effect is more
important than the other for the suppression of the energy gap.
Setting aside how to distinguish the two effects, we may
concentrate on the relationship between the suppression of the
energy gap and high quasiparticle current density injected
directly. In the four-terminal measurements, there are two parts
of quasiparticle current which cause the suppression of the energy
gap. One is from the electrode junctions through which the bias
current passes. A strong quasiparticle current will inject into
the effective junctions from the electrode junctions and cause a
nonequilibrium effect. Besides, the overheating effect in the top
electrode junctions will also cause an increase of the temperature
of IJJs. The second part comes from the measured effective
junctions themselves. The quasiparticle current in one IJJ will be
injected into the adjacent junctions through the common Cu-O
layers, and cause a nonequilibrium effect as well as Joule
heating. Since the sizes of the electrode junctions are smaller
than the effective junctions, the current density in the effective
junctions is smaller with the same current through. Hence, the
quasiparticle current density in the electrode junctions plays a
key role in the suppression of the energy gap, i.e. the size of
the top electrode through which the current passes will be an
important factor for energy gap suppression.

\par
\begin{figure}[tb]
\begin{center}
\includegraphics[width=.6\textwidth]{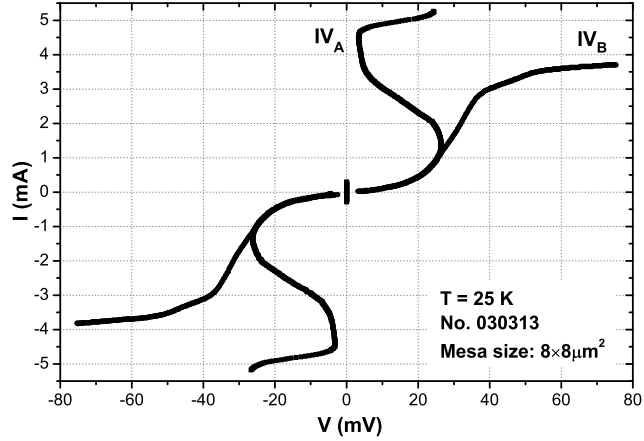}
\end{center}
\caption{$I-V$ curves of a single IJJ when the bias current passes
through different top electrodes. The sizes of the two top
electrodes are $\mathrm{1.5 \times 8 \mu m^2}$ ($E_A$) and
$\mathrm{4.5\times 8 \mu m^2}$ ($E_B$) respectively. $IV_i$
denotes that the bias current passes through top electrode $E_i
(i=A,B)$.} \label{IV2}
\end{figure}

If we design a mesa with two top electrodes of different size, we
should observe different levels of the gap suppression for the
same IJJs when the current passes through the two different top
electrodes. We did observe such a behavior in the following
measurement. For a sample of 3 effective junctions, the sizes of
the two top electrodes were $\mathrm{4 \times 8 \mu m^2}$ ($E_1$)
and $\mathrm{2 \times 8 \mu m^2}$ ($E_2$) in a mesa with size of
$\mathrm{8 \times 8 \mu m^2}$ in the $ab$ plane. When the bias
current is low (less than $5I_c$), the same $I-V$ curves with
three quasiparticle branches were observed with bias current
through the two different top electrodes. However, when the bias
current $I>5I_c$, although the back-bending structure was observed
in both two cases, the suppression of the energy gap was
considerably stronger when the bias current passed through $E_2$
than when it passed through $E_1$. This is consistent with the
current density through the top electrodes. Furthermore, to
minimize the influence of the energy gap suppression from the
effective junctions themselves, we did the similar experiment on a
sample with only one effective IJJ. Two electrodes, of the sizes
$\mathrm{1.5 \times 8 \mu m^2}$ ($E_A$) and $\mathrm{4.5 \times 8
\mu m^2}$ ($E_B$) respectively, are made on the top of the mesa.
The measured $I-V$ curves are shown in Figure \ref{IV2}. When a
bias current passes through $E_A$, the $I-V$ curve showed a
back-bending structure ($V_{gmin}= 0.16\frac{2\Delta}{e}$).
However, when the bias current is through $E_B$, instead of the
back-bending structure, an ordinary quasiparticle branch was
observed with a suppressed gap voltage $V_g/2=16.5 mV< 25 mV$.
This indicates that the suppression of the energy gap still
existed though it seemed not serious in the $I-V$ curve.

\par
\begin{figure}[tb]
\begin{center}
\includegraphics[width=.6\textwidth]{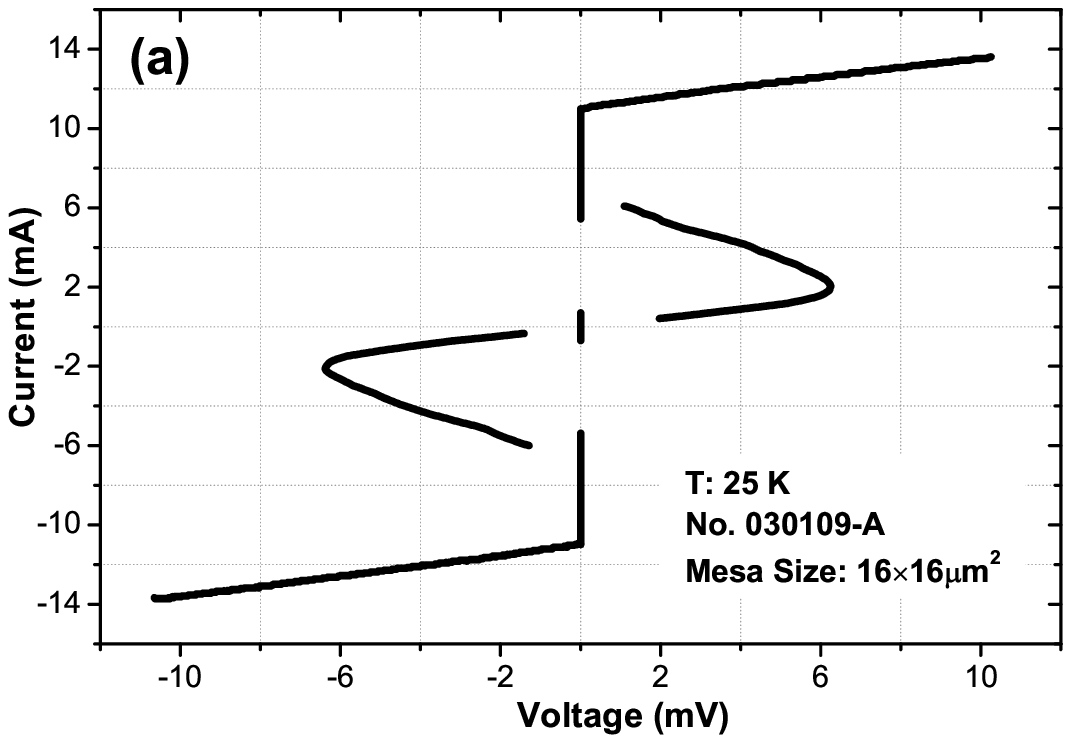}
\includegraphics[width=.6\textwidth]{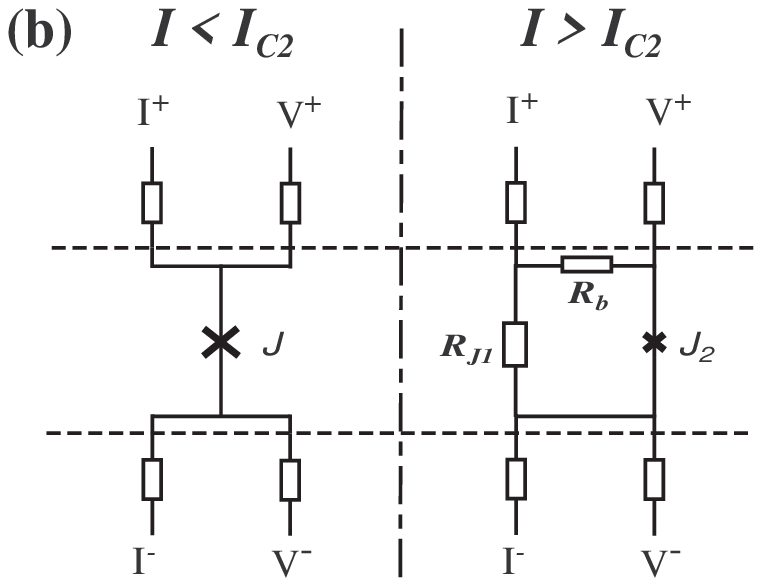}
\end{center}
\caption{(a) Novel $I-V$ curve of a single IJJ. The sizes of two
top electrodes are both $6 \times 16 \mu m^2$; (b) Equivalent
circuits for the single IJJ at high bias current, where $I_{c2}$
denotes the bias current at which the reentrance of super-current
happens.} \label{IV3}
\end{figure}

Most surprisingly, in some samples with a single IJJ, a novel
super-current reentrance was observed instead of the back-bending
structure (shown in Figure \ref{IV3}(a)).  This novel phenomena
was observed in samples of different sizes ($8 \times 8 \mu m^2$
and $16 \times 16 \mu m^2$) and under different measurement
conditions (samples placed in the vacuum chamber of a cryocooler
or directly immersed in Liquid Helium). However, normal
multi-branch $I-V$ curves similar to Figure \ref{IV-3T} was
observed when the sample was measured in three-terminal method.

\par
Such a phenomena can probably be attributed to strong heating
effect. As a matter of fact, with increase of the bias current,
the nonequilibrium effect and Joule heating effect become stronger
and stronger, suppressing the energy gap more and more. When the
bias current reaches $I_{c2}$, at which the reentrance occurs,
further increase of the temperature may cause some Cu-O layers of
the junction to enter the normal state. Regarding the junction we
measured, the part of the upper Cu-O layers of the IJJ under the
top electrode through which the bias current passes will first
enter the normal state due to the poor thermal conductivity.  Then
the junction is no longer a uniform one. The equivalent circuits
are shown in Figure \ref{IV3}(b). In this case, the voltage we
measure is the one across that part of junction ($J_2$) which
still remains in the zero-voltage state; i.e., the voltage is
zero. When the bias current is high enough, the current through
$J_2$ is larger than its critical current, $J_2$ will switch to
non-zero voltage state and the super-current disappears again. For
samples with a few effective junctions, the likeliest part to
reach a high temperature is the upmost Cu-O layers beneath the top
electrode due to the strong heating effect from the top electrode.
Other inner Cu-O layers will have a relatively lower temperature.
That may explain why we can only observe the super-current
reentrance effect in a single IJJ.

\par
Since a quasiparticle current can always pass through the
electrodes and then inject into mesa-structured IJJs, the
suppression of the energy gap is difficult to avoid though
sometimes it is not serious. Especially, when the sample is
fabricated with one top electrode and measured in a 3-terminal
method, we often observed a normal quasiparticle branch with a
higher energy gap similar at almost 25 meV. Nevertheless, a
surface junction is inevitably introduced, which causes series
resistance to occur in the measurement.

\par
The double-sided fabrication technology developed by Wang \etal
provides a feasible way to fabricate IJJs with pure
superconducting contact electrodes and without the influence of
quasiparticle current injected through the electrodes
\cite{Wang:PRL01,Wang:IEICE02}. Yet, a back-bending structure was
still observed due to a large junction number \cite{Wang:PRL01}.
The unsolved problem is how to decrease the junction number to 1
with double-sided technology. With this solution, it may be
possible to realize an ideal $I-V$ curve of IJJ without energy gap
suppression with such a new configuration.

\section{Summary}

In summary, we observed a suppression of the energy gap under high
bias current in mesa-structured IJJs. We have discussed its causes
and measured the variance of gap suppression with temperature.
With the design of different sizes of top electrodes in
mesa-structured IJJs, we observed different levels of energy gap
suppression corresponding to different quasiparticle current
densities through the top electrodes. A novel super-current
reentrance phenomena was observed and discussed. We suppose that
an ideal $I-V$ curve of IJJ with correct energy gap may be
observed in single IJJ fabricated by the double-sided technology.

\ack

The authors acknowledge A. Yurgens, D. Winkler, T. Claeson for
reading and commenting on the manuscript, and thank T. Xiang and
Z. M. Ji for helpful discussions. This work was supported by a
grant from the Major State Basic Research Development Program of
China (No G19990646), the National High-tech Research and
Development Programme of China and Swedish Foundation for
Strategic Research (SSF) through the OXIDE program.

\end{document}